\documentclass[12pt]{aipproc} 
\usepackage{epsf} 
\usepackage{epsfig} 
 
\newcommand{\sss}{\scriptscriptstyle}

\def\bc{\begin{center}}
\def\ec{\end{center}}
\def\be{\begin{equation}}
\def\ee{\end{equation}}
\def\bea{\begin{eqnarray}}
\def\eea{\end{eqnarray}}
\def\nn{\nonumber}

\layoutstyle{6x9}
\begin{document}
\title{TWO BODY $B$ DECAYS, FACTORIZATION AND 
$\Lambda_{QCD}/m_{b}$ CORRECTIONS\thanks{Talk given by G. Martinelli}}
\author{M.~Ciuchini}{
  address={Dip. di Fisica, Univ. di Roma III and INFN,
           Sezione di Roma Tre, Via della Vasca Navale 84, I-00146   
           Roma, Italy},
  email={},
  thanks={}
  }
\author{E.~Franco}{
  address={Dip. di Fisica, Univ. ``La Sapienza'' and INFN,
           Sezione di Roma, P.le A. Moro, I-00185 Rome, Italy},
  email={},
  thanks={}
  }
\author{G.~Martinelli}{
  address={Dip. di Fisica, Univ. ``La Sapienza'' and INFN,
           Sezione di Roma, P.le A. Moro, I-00185 Rome, Italy},
  email={},
  thanks={}
  }
\author{M.~Pierini}{
  address={Dip. di Fisica, Univ. ``La Sapienza'' and INFN,
           Sezione di Roma, P.le A. Moro, I-00185 Rome, Italy},
  email={},
  thanks={}
  }
\author{L.~Silvestrini}{
  address={Dip. di Fisica, Univ. ``La Sapienza'' and INFN,
           Sezione di Roma, P.le A. Moro, I-00185 Rome, Italy},
  email={},
  thanks={}
  }
\begin{abstract}
By using the recent experimental measurements
of $B \to \pi\pi$ and $B \to K \pi$ branching ratios, we find that
the amplitudes computed at the leading order of the $\Lambda_{QCD}/m_{b}$
expansion  disagree with  the observed $BR$s,  even taking
into account the uncertainties of the input parameters. Beyond the 
leading order,  Charming and GIM penguins allow to reconcile the theoretical predictions with the
data.  Because of these large effects, we conclude, however, that it
is not possible, with the present theoretical and experimental
accuracy, to determine the CP violation angle $\gamma$ from these
decays.  We compare our results with those obtained with the 
parametrization of the chirally enhanced non-perturbative 
contributions  by BBNS. We also predict large asymmetries for
several of the particle--antiparticle $BR$s, in particular $BR(B^{+}
\to K^+ \pi^0) $, $BR(B_d \to K^+ \pi^-) $ and $BR(B_d \to \pi^+
\pi^-)$. 
\end{abstract}
\maketitle
 
\section{Introduction} \label{intro}
The advent of the $B$ factories, Babar~\cite{babar} and 
Belle~\cite{belle}, opens new  perspectives for very precise measurements of non-leptonic 
$B$-decays~\cite{ferroni} and  calls for a significant improvement of
the theoretical predictions.  In this respect,  important progress
has been recently achieved by systematic studies of factorization  
in  $B$
decays~\cite{li,beneke}. These studies  confirmed the physical idea~\cite{previous} that factorization holds
in heavy  hadron decays,  $m_{Q} \gg \Lambda_{QCD}$, for 
the leading terms of the $\Lambda_{QCD}/m_{Q}$ expansion. They  
leave open, however, the central question of whether  i) the leading terms  predict  with  
sufficient accuracy the relevant $B$-meson decay rates or ii) the power-suppressed 
corrections, which cannot be evaluated without some model assumption, 
are phenomenologically important.  This problem has been  addressed in a 
series of papers~\cite{charming}--\cite{lorolast}. In particular, the main
conclusion of ref.~\cite{noilast} is that {\it non-perturbative}
$\Lambda_{QCD}/m_{b}$  corrections from  the leading operators of the effective weak
Hamiltonian, conventionally called Charming (or GIM) penguins,  are very important in cases where the
factorized amplitudes are either colour or Cabibbo suppressed.   One 
of the  consequences of this observation is that  factorization at 
the leading order in $\Lambda_{QCD}/m_{b}$ is unable to reproduce the observed 
 $B \to \pi\pi$ and $B \to K \pi$ $BR$s even taking into account the uncertainties of the input parameters.
 
 The results of ref.~\cite{noilast}  seem to be at variance with 
the conclusions of ref.~\cite{lorolast} where it is stated 
that, in the QCD factorization approach, ``an acceptable fit to the branching 
fractions is obtained even if we 
impose $\gamma < 90^{o}$ as implied by the standard constraints on  the 
unitarity triangle''~\footnote{ Indeed the analysis of 
ref.~\cite{Ciuchini:2000de}  gives  $\gamma = (54.8 \pm 6.2)^{o}$ and 
ref.~\cite{hocker} quotes  $34^{o} < \gamma < 82^{o}$.}. 

Given  the contradictory conclusions of the different studies, it is   
very important to clarify the situation by comparing input parameters,
methods of analysis and results.  In order to help the 
debate on this subject, we discuss in the following:
\begin{itemize} 
\item   The theoretical framework of  factorization and  the calculation 
of the amplitudes at the leading order  of the $\Lambda_{QCD}/m_{b}$  
expansion;
\item A  comparison of the leading terms, including their 
uncertainties,  with the measured  $B \to \pi\pi$ and $B \to K \pi$ branching ratios;
\item  Results for the $B \to \pi\pi$ and $B \to K \pi$ $BR$s  including 
 charming (and GIM) penguins;
\item   A comparison of the results obtained with  charming and GIM penguins~\cite{noilast} 
with the model of the  $\Lambda_{QCD}/m_{b}$ corrections 
adopted  in  \cite{lorolast}.
\end{itemize}
\section{Factorization}
\label{sec:factorization}

In this section we recall  the basic ingredients necessary to compute 
the relevant amplitudes  either whithin factorization or  including  the corrections 
which arise at higher order in the  $\Lambda_{QCD}/m_{b}$  expansion.  

The physical amplitudes for $B \to K \pi$ and $B \to \pi \pi$ can be 
conveniently written in terms of RG-invariant parameters built using
the Wick contractions of the effective Hamiltonian~\cite{BS}.  In the
heavy quark limit, following the approach of ref.~\cite{beneke}, it is
possible to compute these RG invariant parameters using factorization.
The formalism of ref.~\cite{beneke}  has been developed so that it is 
also possible to include the perturbative corrections to order $\alpha_{s}$, i.e. at the
next-to-leading order in perturbation theory~\footnote{ An alternative
framework,  provided by the approach of ref.~\cite{li}, will not be 
discussed here. For a recent discussion see also~\cite{lastchris}.}.

For the sake of discussion, it is instructive to start from the explicit expression of the $B_d \to K^+ \pi^-$
amplitude. In terms of the parameters defined in~\cite{BS}, this amplitude reads 
\begin{eqnarray} {\cal
    A}(B_d \to K^+ \pi^-) = &-& V_{us} V_{ub}^*
  \Bigl(E_1(s,u,u;B_d,K^+,\pi^-) - P_1^{\sss {\rm
      GIM}}(s,u;B_d,K^+,\pi^-)\Bigr)\nn \\ 
&+& V_{ts} V_{tb}^*
  \,P_1(s,u;B_d,K^+,\pi^-)\,.  \label{eq:uno}
\end{eqnarray} 
We have 
\begin{eqnarray}
  E_1(s,u,u;B_d,K^+,\pi^-)&=& a_1^u(K \pi) \langle Q_1^u\rangle_{\rm
    fact} + a_2^u(K \pi) \langle
  Q_2^u\rangle_{\rm fact} + \tilde E_1\nn \\
  P_1(s,u;B_d,K^+,\pi^-)&=&\sum_{i=3}^{10} a_i^c(K\pi) \langle Q_i
  \rangle_{\rm fact} +\tilde P_1 \nn \\ 
P_1^{\sss {\rm
      GIM}}(s,u;B_d,K^+,\pi^-)&=& \sum_{i=3}^{10}
  (a_i^c(K\pi)-a_i^u(K\pi)) \langle Q_i \rangle_{\rm fact} +\tilde
  P_1^{\sss {\rm GIM}}\, ,   \label{eq:due}
\end{eqnarray} 
 where $\langle Q_i \rangle_{\rm fact}$ denotes the factorized matrix 
 elements and $a^{f}_i$ the  parameters  introduced   in~\cite{beneke}. 
Eqs.~(\ref{eq:uno}) and ~(\ref{eq:due}) are exact. Unfortunately, similar equations
 require the knowledge of several non-perturbative parameters, 
which at present cannot be extracted from the data.  To be 
predictive, we will then use our
physical intuition  to reduce their number. 
In eqs.~(\ref{eq:uno}) and (\ref{eq:due}):
\begin{enumerate} \item  The CKM matrix elements 
can either be taken from other experimental measurements or from a 
fit to the non-leptonic $BR$s, assuming that factorization is 
accurate enough;  
\item The coefficients $a_i^f(M_{1} M_{2})$ (e.g. $a_1^u(K 
\pi)$) are renormalization group invariant, as the corresponding  
factorized matrix elements, and have been computed  perturbatively  at 
the NLO in refs.~\cite{beneke,lorolast};
\item  The coefficients $a^{f}_{6}$ and 
$a^{f}_{8}$, which have also been computed  to one-loop order,  are instead 
scheme dependent. Their scheme-dependence is cancelled by the 
hadronic matrix elements of the penguin operators $Q_{6}$ and $Q_{8}$ 
respectively. 
Assuming  factorization for the  chirally enhanced 
contributions~\cite{beneke}, the latter can be expressed in terms of the ratio
\begin{equation} r^{K}_{\chi}(\mu) = \frac{2 m_{K}^{2}}{\overline{m}_{b}(\mu) 
(\overline{m}_{s}(\mu)  +\overline{m}_{q}(\mu)  )}\, ,  \end{equation} 
which is formally of ${\cal O}(\Lambda_{QCD}/m_{b})$ but numerically 
important (an analogous parameter $r^{\pi}_{\chi}(\mu)$ can be defined 
for $\pi\pi$ decays).  We will discuss  the r\^ole of these terms 
below.
\item The leading amplitudes $\langle Q_i \rangle_{\rm fact}$ are
computed in terms of decay constants and semi-leptonic form factors.  
The form factors can either be taken from theoretical 
calculations~\cite{latticeff,qcdsrff} or fitted from the experimental 
$BR$s (the possibility  of extracting them from the corresponding 
$B \to \pi$  semileptonic  form factor at 
small momentum transfer is at present rather remote).  
\item  The tilded parameters, namely 
$\tilde P_1$, $\tilde   P_1^{\sss {\rm GIM}}$ and $\tilde E_1$,   are genuine, non-perturbative $\Lambda_{QCD}/m_{b}$ 
corrections which cannot  be computed at present.  
If we neglect Zweig-suppressed contributions,  by $SU(2)$ symmetry one can show that all the
Cabibbo-enhanced $\Lambda_{QCD}/m_{b}$ corrections to $B \to K \pi$
decays can be reabsorbed in  $\tilde P_{1}$.
Several corrections are contained in $\tilde P_{1}$: this parameter
includes not only the charming penguin contributions, but also
annihilation and penguin contractions of penguin operators. It does
not include leading emission amplitudes of penguin operators
($Q_3$--$Q_{10}$) which have been explicitly evaluated using
factorization.  Had we included these terms, this contribution would
exactly correspond to the parameter $P_1$ of ref.~\cite{BS}.  The
parameter $\tilde P_{1}$ ($P_{1}$) encodes automatically not only the
effect of the annihilation diagrams considered in~\cite{keum}, but
all the other contributions of ${\cal O}(\Lambda_{QCD}/m_{b})$ with
the same quantum numbers of the charming penguins. In this respect it is
the most general parameterization of all the perturbative and
non-perturbative contributions of the operators $Q_{5}$ and $Q_{6}$
($Q_{3}$ and $Q_{4}$), including the worrying higher-twist infrared
divergent contribution to annihilation discussed in
ref.~\cite{lorolast,beneke2}.  The parameter $\tilde P_{1}$  has
the same quantum numbers and physical effects
as the original charming penguins proposed in~\cite{charming}, 
although it has a more general meaning.  
\item If one also includes $B \to \pi\pi$ decays we have several other
parameters, for example $P_{1}^{\sss {\rm GIM}}$ and $P_{3}$, in the formalism of
ref.~\cite{BS}. A closer look to $P_{3}$ shows that this term is due
either to Zweig suppressed annihilation diagrams (called CPA and DPA
in ref.~\cite{charming}) or to annihilation diagrams which are colour
suppressed with respect to those entering $\tilde P_{1}$. For
this reason in ref.~\cite{noilast}  $P_{3}$ was taken to be zero.  $\tilde 
P_1$ is equal to the corresponding parameter in $K \pi$ decays if $SU(3)$
symmetry is assumed.   In our analysis we have used the same value of $\tilde
P_1$ for all $K \pi$ and $\pi \pi$ channels. 
$P_{1}^{\sss {\rm GIM}}$ will be
discussed later on.
\end{enumerate}
 
We are now ready to discuss {\it  factorization} for the 
leading terms of the $\Lambda_{QCD}/m_{b}$ expansion. 
Factorization is the theory of non leptonic decays which is obtained in the limit $m_{b} 
\to \infty$.  Thus it consists  in neglecting {\it all}  terms of ${\cal 
O}(\Lambda_{QCD}/m_{b})$ ($\tilde E_{1}=0$, $\tilde P_{1}=0$, 
$r^{K,\pi}_{\chi}=0$, etc.).
 At   lowest order in perturbation theory,  called also na\"{\i}ve 
factorization,  the $a^{f}_{i}$ are simple combinations of the Wilson 
coefficients  and do not depend on the hadron wave functions. 
The inconvenience of na\"{\i}ve  factorization is that  physical amplitudes 
still have a marked dependence on the renormalization scale because, 
contrary to  the Wilson coefficients,  the factorized matrix elements are 
scale independent. 

The scale dependence is reduced by working at 
${\cal O}(\alpha_{s})$, both for the Wilson coefficients and the 
matrix elements.  In refs.~\cite{beneke} it has been shown that, at 
this order,  all the  dangerous infrared divergences can be reabsorbed in the definition of 
the hadronic wave functions. For the  leading terms in the  
$\Lambda_{QCD}/m_{b}$ expansion there are strong arguments to support 
the idea that   this will remain true    at all orders in 
$\alpha_{s}$, see also ref.~\cite{allorders}.  Thus, in the limit 
$m_{b} \to \infty$, it is likely that factorization    is  preserved by strong 
interactions. 
At ${\cal O}(\alpha_{s})$ or higher, 
the coefficients  $a^{f}_{i}$ depend  on the specific detail of the hadron 
wave-fuctions. For this reason, the uncertainties 
relative to the wave functions, as the residual renormalization scale  dependence, must be taken into account
in the evaluation of the  uncertainties for the theoretical 
predictions.   The approximation in which we neglect all the 
$\Lambda_{QCD}/m_{b}$ corrections, but include the perturbative 
corrections to the leading contribution,  is called  QCD factorization 
or simply  factorization. Factorization implies an 
important consequence:   predictions of   non-leptonic decay 
rates are {\it model independent} to the extent that  the few relevant hadronic 
parameters, namely the kaon and pion decay constants, $f_{K,\pi}$, the 
semileptonic form factors, $f_{K,\pi}(0)$, and  the hadronic wave functions are
known.

In cases like $B \to K \pi$ decays, where the factorized amplitudes are 
Cabibbo suppressed,   the corrections of ${\cal O}(\Lambda_{QCD}/m_{b})$, which unfortunately are 
{\it model dependent},  become  important.  At lowest order in $\alpha_{s}$, the chirally enhanced terms proportional 
to  $r^{K}_{\chi}$ ($r^{\pi}_{\chi}$)  are  computable  by  assuming  
that factorization can be applied beyond the leading order.  
A   substantial  difficulty arises,  however, at  ${\cal 
O}(\alpha_{s}\Lambda_{QCD}/m_{b})$. Although the chirally  enhanced corrections from $Q_{6,8}$ are infrared finite, 
other  contributions of the same order from different operators are infrared 
divergent, signaling  that they  belong to the class of the non-perturbative    contributions 
which appear beyond factorization. These 
cannot be predicted using the same hadronic quantities  of the  factorized amplitudes. For this 
reason, any phenomenological analysis which  aims at including in a coeherent way  
the terms of ${\cal O}(\alpha_{s}\Lambda_{QCD}/m_{b})$  is forced to introduce 
extra model-dependent  non-perturbative   parameters besides 
$f_{K,\pi}$, $f_{K,\pi}(0)$ and the hadronic wave functions.   This implies   
that, {\it at  ${\cal 
O}(\Lambda_{QCD}/m_{b})$,   model dependence is unavoidable} (even in 
the subsector of the chirally enhanced contributions) and it is present 
in both the analyses  of ref.~\cite{noilast}  and ref.~\cite{lorolast},  
which we will compare below.  

Different model-dependent assumptions were made in the two approaches:
\begin{enumerate} \item  In ref.~\cite{noilast}, $\tilde E_{1}$ and 
$\tilde P_{3}$ were neglected and $SU(2)$ symmetry was assumed for  $\tilde 
P_{1}$   ($SU(3)$ when it was used for $B \to  \pi\pi$ decays). The 
same approximations were made  for $\tilde P_1^{\sss {\rm GIM}}$.  The 
complex parameter $\tilde P_{1}$ ($\tilde P_1^{\sss {\rm GIM}}$) was then fitted to reproduce the 
experimental $BR$s. 
\item In ref.~\cite{lorolast} the  effects of the chirally enhanced $\Lambda_{QCD}/m_{b}$ 
corrections were either computed 
perturbatively or encoded in two complex phenomenological parameters  
called $X_{H}$ and $X_{A}$. An uncertainty of 100 $\%$ to the ``default'' 
values (e.g. $X_{H}=2.4$) was assigned  to these parameters in order 
to determine allowed bands  for the predicted $BR$s. The bands include 
all other sources of uncertainties. 
\end{enumerate}
 
We now compare the leading amplitudes to the experimental results in 
order  to {\it test factorization}. 
Before using predictions based on factorization to test the Standard 
Model and look for signals of new physics, it is crucial to check how large are the 
errors induced by our ignorance of the ${\cal 
O}(\Lambda_{QCD}/m_{b})$ corrections which we are unable to compute.
Our position, indeed, is that we have more confidence in the SM rather than 
in factorization.  In order to test factorization, we ought to use all 
the information that we have from other measurements. Thus, for 
example, whereas 
the size of the error on $\vert V_{cb}\vert$ and $\vert V_{ub}\vert$ can 
be  debated, there is no question that these experimental inputs must 
be included in any analysis that aims at testing (or using) factorization.  We 
 also stress that the value of $\vert V_{ub}\vert$ is not expected to be 
affected by the presence of new physics beyond the SM.

The CP parameter $\gamma$ does in general change if  there is 
physics beyond the SM. It remains an interesting exercise, however,  to verify 
whether, by taking the value of $\gamma$ from the Unitarity Triangle 
Analysis (UTA) in the SM,  the predicted $BR$s are in agreement with 
the data. If, by  using $\gamma$ from the UTA, one is unable to reproduce 
the experimental  $B \to \pi\pi$ and $B \to K \pi$  $BR$s, this implies that ``either
there is new physics or $\Lambda_{QCD}/m_{b}$ corrections are
important''~\cite{noilast}. 

In our analysis we have used the likelihood method which has been 
described in all details in ref.~\cite{Ciuchini:2000de}. Without 
entering in the ``ideological'' controversy about frequentistic and 
bayesian methods, we only note here that in~\cite{Ciuchini:2000de} it 
has been shown that, at 95 \% C.L.,  the Bayesian analysis give the same results as 
the frequentistic Babar Scanning method (and its variations) when the 
same inputs are used. Thus we will present our contour plots, 
corresponding to fig.~17  of ref.~\cite{lorolast}, both with 
factorization plus   chirally enhanced contributions
and with the non-factorizable charming (and GIM) penguin 
corrections.  
Besides this, we will also give tables with the relevant $BR$s, both 
in the factorization approximation  with chirally enhanced terms  and with the charming (and GIM) penguin 
corrections included.  

We end this section with some  remarks.  In our approach,  we 
have first  checked that, within factorization and the SM, it is impossible to 
fit the experimental $BR$s.   The  $\Lambda_{QCD}/m_{b}$ terms, that 
we are then  forced to include  in order to reproduce the experimental results, are 
non-perturbative  quantities, infrared divergent in perturbation 
theory, on which we do not have any knowledge {\it a priori}. For this reason we 
decided to fit them on the data.  The experimental numbers are nicely 
reproduced and  the corrections to  factorization  are well 
consistent with the expected size (i.e.  $\tilde P_{1}$ is of ${\cal 
O}(\Lambda_{QCD}/m_{b})$ with respect to the leading contributions).

In ref.~\cite{lorolast}  the subleading power corrections 
are varied in  predefined intervals and the change in the predicted 
$BR$s is interpreted as uncertainty on the factorized   amplitudes.
In our opinion the uncertanties on factorization are only those coming from 
the CKM matrix elements or the form factors, etc. The $\Lambda_{QCD}/m_{b}$ terms 
instead are really contributions beyond factorization: if they are 
necessary to reproduce the data then it is not possible to make model 
independent predictions. This would remain true even if we knew without any uncertainty the 
hadronic parameters entering at the lowest order of the 
$\Lambda_{QCD}/m_{b}$ expansion~\footnote{ Note that some $BR$s are dominated by 
the $\Lambda_{QCD}/m_{b}$ corrections.}.  

Thus we are in the Bermuda triangle:
i) without the ${\cal  O}(r_{\chi})$  and ${\cal  O}(\alpha_{s}r_{\chi})$ terms, that is within 
QCD factorization,  we cannot reproduce the data; ii) the  
inclusion of the computable  subset  of  ${\cal  O}(\alpha_{s}r_{\chi})$  
terms  only is inconsistent since there is no reason to exclude 
the other  non-perturbative non-computable contributions of the same order.
In any case, we will show that, by using all the available experimental
information, also this case is very difficult to reconcile with the 
data; iii)  the complete 
set of  ${\cal  O}(\alpha_{s}r_{\chi})$  corrections leads us beyond factorization 
and the results are model dependent.  Indeed there is no  proof that 
the one-loop   finite chirally enhanced terms remain infrared finite 
at higher orders in $\alpha_{s}$. Moreover, if the corrections of
${\cal  O}(\alpha_{s}r_{\chi}\sim \alpha_{s} \Lambda_{QCD}/m_{b})$ 
are phenomenologicaly important, it is 
difficult to understand why, at the same level of numerical accuracy,  other non-chirally 
enhanced non-perturbative $\Lambda_{QCD}/m_{b}$ terms  should not  also be taken into account. 

The sad conclusion is that the very nice theory of QCD factorization 
developed in~\cite{beneke} is insufficient to fit the data because power 
corrections, which are model dependent, are important in $B\to K\pi$ 
and $B\to\pi\pi$ decays.   
Finally, model dependence does not implies that we are unable to make 
any prediction. If the assumptions made in our approach are 
reasonable~\footnote{ For example that 
we may neglect $\tilde E_{1}$.},  by fitting $\tilde  P_{1}$ to the data and with the increasing 
experimental precision we may hope to extract also the value of $\gamma$ 
or to constrain $\sin 2 \alpha$. 
\section{Results}
\label{sec:results}

In this section we  present our analysis and a detailed comparison 
with the results of ref.~\cite{lorolast}. 
In order to obtain our results we used the likelihood method 
as in~\cite{noilast}. The input parameters, given in table~\ref{tab:inputs},  
are also the same but for a few  differences:
\begin{itemize} 
\item We added the ${\cal O}(\alpha_{s})$ corrections to the 
coefficients of the penguin and electropenguin operators computed 
in~\cite{lorolast},  which appeared after the completion of~\cite{noilast}.  
In this respect the criticisms of ref.~\cite{beuc}
do not apply to the present analysis. We will see below that, 
even including these new ingredients, 
the main physics conclusions of ref.~\cite{noilast} are confirmed.

\item As for the matrix elements of $Q_{6,8}$, we include 
the ``computable'' factorizable chirally enhanced terms in the 
definition of $\tilde  P_{1}$, which in this way contains all the 
possible Cabibbo enhanced, model-dependent  corrections of ${\cal 
O}(\Lambda_{QCD}/m_{b})$. In practice this corresponds to fit $\tilde 
P_{1}$ as in ref.~\cite{noilast} with $r_{\chi}^{K,\pi}=0$. There is no
substantial difference between the old choice and the new one,
since the contribution of the chirally enhanced terms and
of $\tilde P_1$ have exactly the same quantum numbers.  
\item Although this is a minor source of uncertainty we also allow a 
variation of the renormalization scale and  of the 
parameters of the hadronic wave  functions.
\end{itemize}
We have also verified that by using the inputs of ref.~\cite{lorolast}
we obtain essentially the same results, and hence arrive to the same physics
conclusions.

In this section we present:
\begin{enumerate}
\item  A brief discussion of  the results obtained in QCD 
factorization, namely with all the terms of ${\cal 
O}(\Lambda_{QCD}/m_{b})$  set to zero. 
\item  The results obtained in our 
approach by fitting $\tilde P_{1}$ and using $\gamma$ as determined 
from the Unitary Triangle Analysis in ref.~\cite{Ciuchini:2000de}
\begin{equation} \gamma = (54.8 \pm 6.2)^{o} \label{eq:gamma} \, .
\end{equation} 
On the basis of this study we predict the $B$--$\bar B$ asymmetries of 
the $BR$s, such as 
\begin{equation}
  {\cal A}(B_d \to \pi^+ \pi^-) = \frac{BR(\bar B^{0}_{d} \to
    \pi^{+}\pi^{-})-BR(B^{0}_{d} \to \pi^{+}\pi^{-})}{BR(\bar
    B^{0}_{d} \to \pi^{+}\pi^{-})+BR( B^{0}_{d} \to \pi^{+}\pi^{-})}
  \, ;
\label{eq:asy} 
\end{equation} 
\item  The results obtained by letting $\gamma$ free and 
a comparison of  our results with those of ref.~\cite{lorolast}.
\end{enumerate}
\subsection*{Results in QCD factorization}
In this subsection we compare the model independent results obtained 
with QCD factorization, namely including only the terms which survive 
when $m_{b}\to \infty$, with the experimental data.  In this case there 
is still  a residual model 
dependence due to our ignorance of the semileptonic form factors 
$f_{K,\pi}(0)$ and, at order $\alpha_{s}$, to the ignorance of the 
hadron wave functions.  We vary the semileptonic form factors with 
flat p.d.f.  in the intervals given in table \ref{tab:inputs} and the 
parameters of the
hadron wave functions in the intervals given  in table 2 of ref.~\cite{lorolast}. 
We also vary the renormalization scale between $m_{b}/2$ and $2 
m_{b}$, see table 1 of ~\cite{lorolast}.  
The data show a generalized disagreement  with the  QCD factorization 
predictions.  In 
particular the allowed region in the $\rho$--$\eta$ 
plane and  the value of $\gamma$  do not have any overlap with 
the corresponding ones from the unitarity triangle analysis~\cite{Ciuchini:2000de}. This 
remains true even if we double the uncertainty on $\vert V_{ub} 
\vert$. We conclude that QCD factorization cannot be reconciled with 
data. 

\begin{table}
\centering
\begin{tabular}{|c|c|c|c|}
\hline
$f_\pi(0)$& $0.27 \pm 0.08$ &
$f_K(0)/f_\pi(0)$ & $1.2 \pm 0.1$ \\ \hline
$\rho $ & $0.224 \pm 0.038$ & $\eta$ & $0.317 \pm 0.040 $ \\ \hline
$BR(B_d \to K^0 \pi^0)$ & $10.3 \pm 2.6 $ &  
$BR(B^+ \to K^+ \pi^0)$ & $12.0 \pm 1.7 $ \\ \hline
$BR(B^+ \to K^0 \pi^+)$ & $17.4 \pm 2.6$ &  
$BR(B_d \to K^+ \pi^-) $& $17.3 \pm 1.6$ \\ \hline
$BR(B_d \to \pi^+ \pi^-) $& $4.4 \pm 0.9 $ &  
$BR(B^+ \to \pi^+ \pi^0) $& $5.3 \pm 1.7$ \\
\hline
\end{tabular}
\caption{Input values used in the numerical analysis. The form
    factors are taken from  
refs.~\cite{latticeff,qcdsrff}, the CKM parameters from
ref.~\cite{Ciuchini:2000de} and the BRs  
correspond to our average of CLEO, BaBar and Belle results 
\cite{cleobr,babarbr,bellebr}.  All the $BR$s are given in units of
$10^{-6}$.} 
\label{tab:inputs}
\end{table}
\begin{table} 
\centering 
\begin{tabular}{|c|c|c|c|c|c|c|c|}
\hline $BR$ & Charming & chirally & BBNS & $BR$ & Charming & 
chirally & BBNS \\ 
& + GIM & enhanced & & &  + GIM & enhanced &\\
 \hline $K^0 \pi^0$ & $8.6 \pm 0.9 $&$3.6
\pm 1.5$ & $4.1 \pm 1.8$
& $K^+ \pi^0$ & $9.8 \pm 1.0 $&$5.7\pm 2.3$ &$6.2 \pm 2.6$
\\ $K^0 \pi^+$ & $18.7 \pm 1.6 $ &$10.2 \pm 4.2$& $10.9 \pm 4.8$ &
 $K^+ \pi^-$ & $17.9 \pm 1.4 $&$8.2\pm 3.4$ &$9.2 \pm 4.0$
 \\ $
\pi^+ \pi^-$ & $4.9 \pm 0.8 $&$9.2\pm 3.8$ & $9.2 \pm 3.8$&
 $\pi^+ \pi^0$ & $3.5 \pm 0.8 $&$5.7 \pm 2.2$  & $6.5 \pm 2.5$
 \\ $ \pi^0 \pi^0$
& $0.6 \pm 0.2 $&$0.2 \pm 0.1$& $0.4 \pm 0.3$ && & &\\ \hline
\end{tabular} 
\caption{$BR$s  with Charming ad GIM penguins,  
with  QCD factorization and the infrared-finite chirally enhanced corrections 
only and with the BBNS model which includes $X_{A,H}$.
 All the $BR$s are given in units of $10^{-6}$.} 
\label{tab:one}
\end{table} 
\subsection*{Factorization with Charming and GIM penguins}
We now discuss the effects of charming  and GIM penguins, parameterized by
$\tilde P_{1}$ and $\tilde P_1^{\sss {\rm GIM}}$.  $\tilde P_{1}$ is a complex amplitude that we fit on
the $B \to K \pi$ $BR$s.  In order to have a reference scale for its
size, we introduce a suitable ``Bag'' parameter, $\hat B_{1}$, by
writing 
\begin{equation} 
\tilde P_{1} = \frac{G_{F}}{\sqrt{2}} \,
  f_{\pi} \, f_{\pi}(0) \, g_{1} \hat  B_{1} \, , 
\end{equation}
where $G_{F}$ is the Fermi constant, $g_{1}$ is a
Clebsh-Gordan parameter depending on the final $K \pi$ ($\pi\pi$)
channel and $B_{1}= \vert B_{1} \vert 
\exp(i \phi )$.  Note that $\hat B_{1}$ differs 
from the parameter defined in ref.~\cite{noilast} because it now 
includes all the chirally enhanced $\Lambda_{QCD}/m_{b}$ corrections, 
part of which were previously explicitly calculated using 
factorization. In a similar way we introduce $\hat B^{\sss {\rm 
GIM}}_{1}$.

\begin{figure}[t]
\centering
\begin{tabular}{cc}
\epsfig{figure=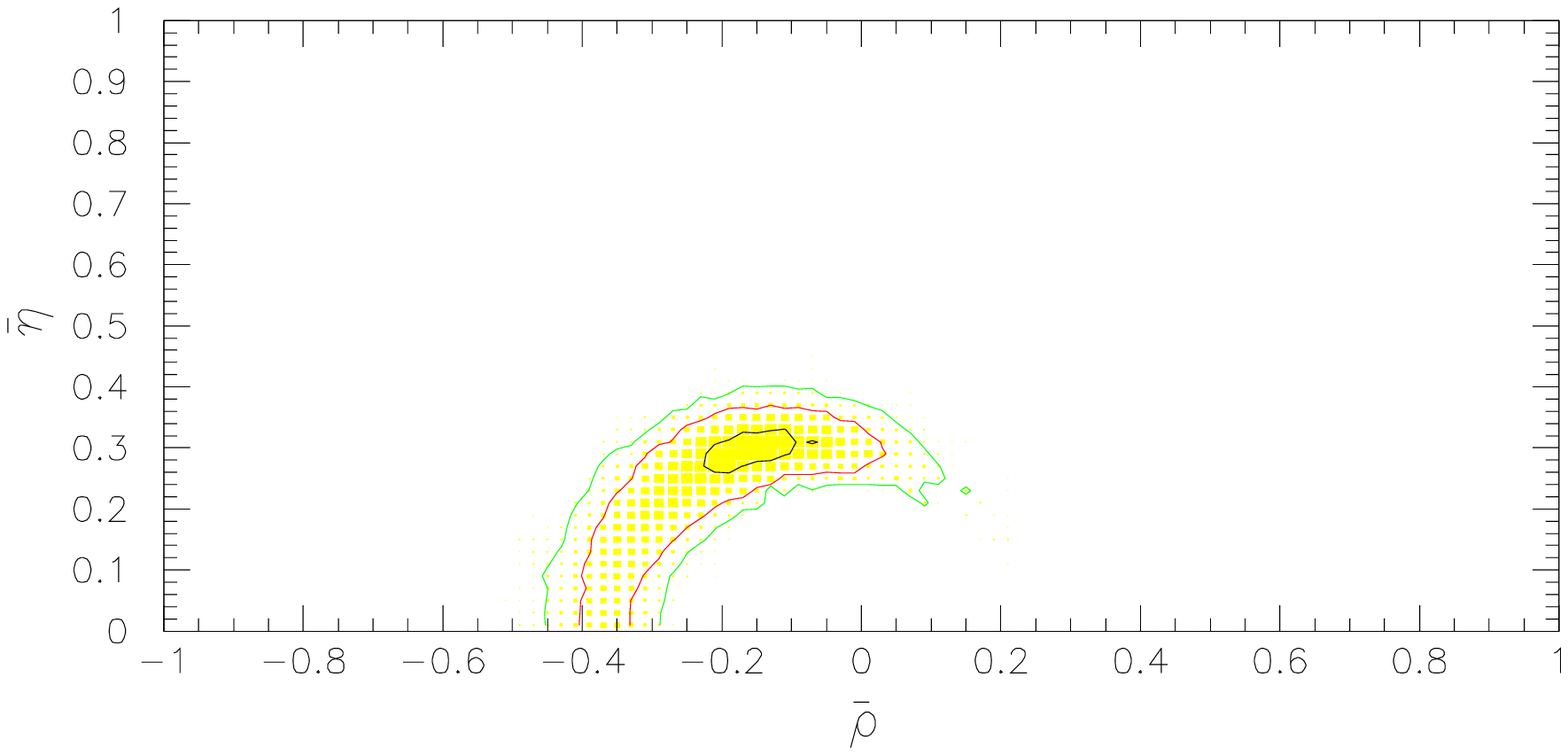,width=3.in} & 
\epsfig{figure=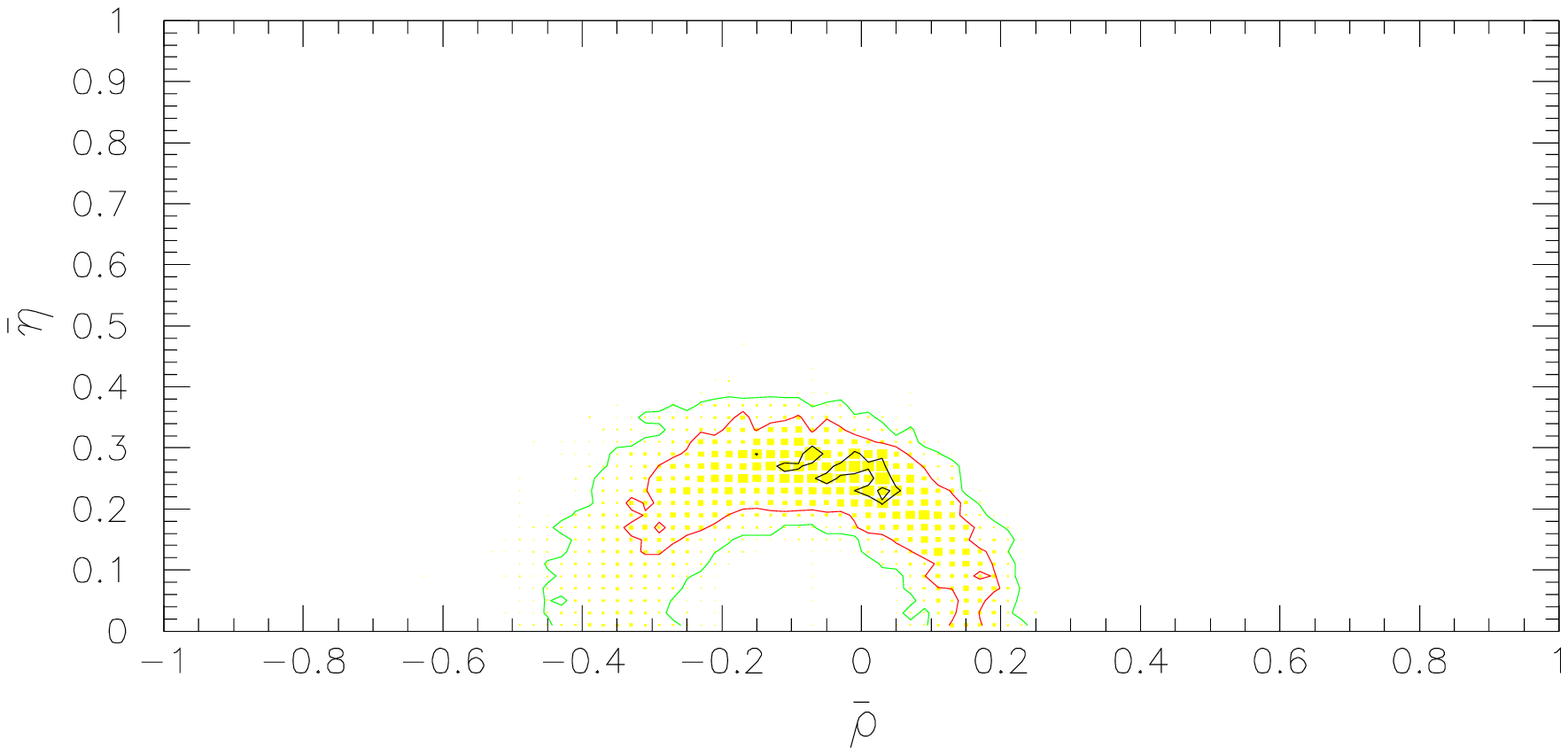,width=3.in} \\ \epsfig{figure=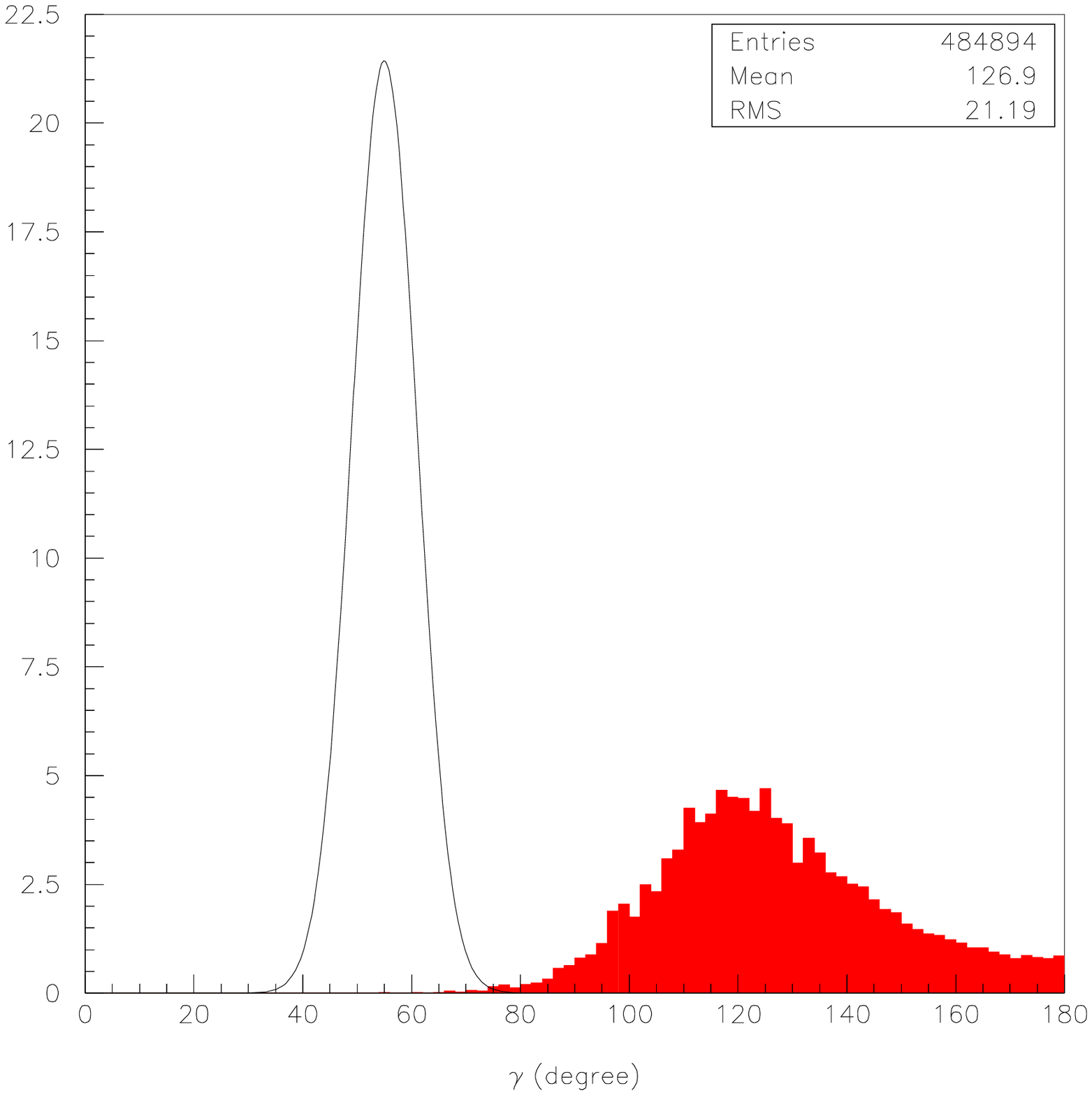,
width=3.in,height=2.8in} & \epsfig{figure=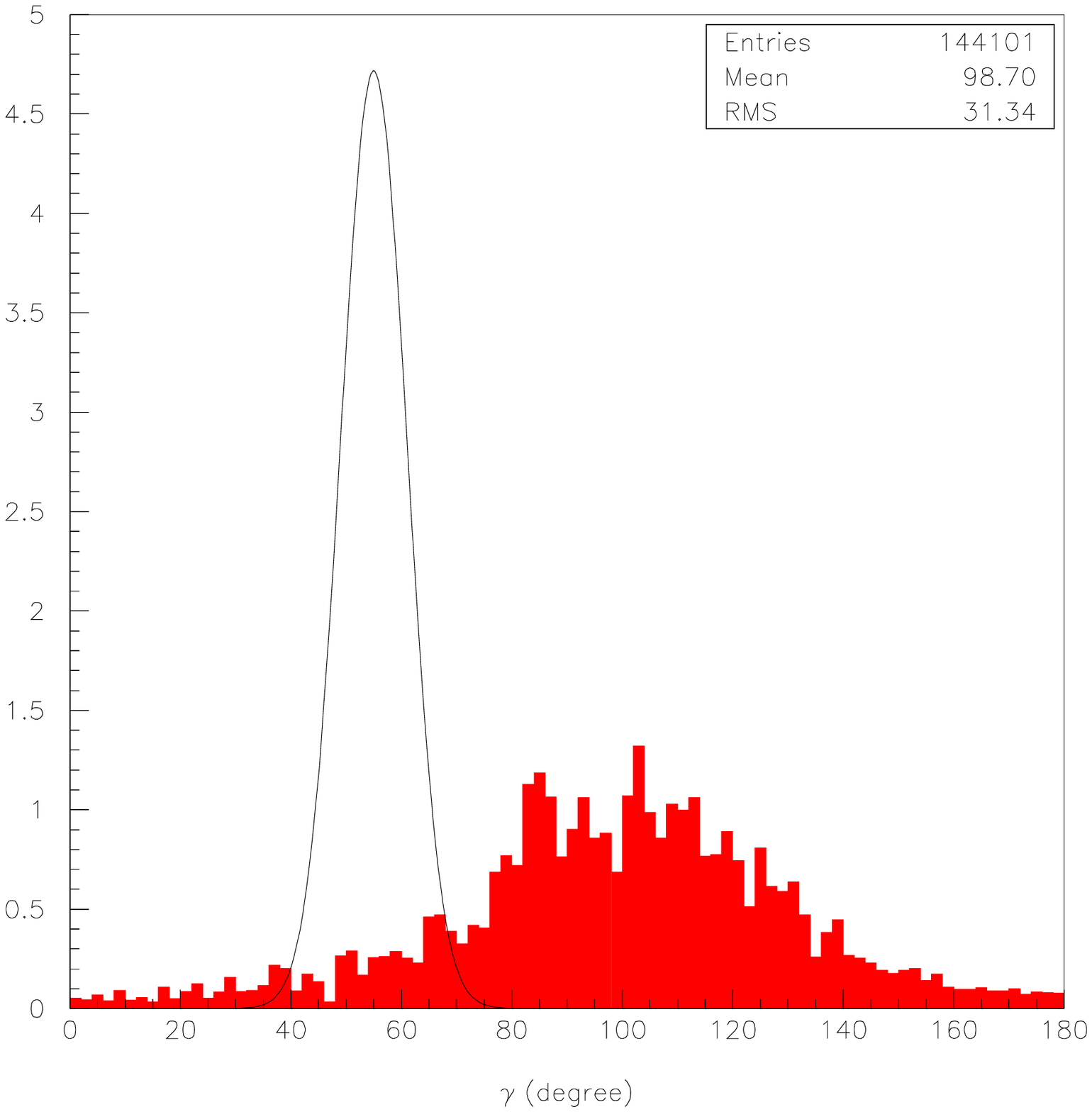,width=3.in,height=2.8in}
\end{tabular}
\caption{$\rho$--$\eta$ contour plots obtained with factorization and
infrared-finite chirally enhanced terms using $\vert V_{ub}\vert$ 
(up-left) or letting $\rho$ and $\eta$ free (up-right).  
We also show a comparison of the p.d.f. of $\gamma$ with the one from
 the UTA analysis of ref.~\protect\cite{Ciuchini:2000de} 
in the two cases.}
\label{fig:lorolastnostro}
\end{figure}

We  fit all the $BR$s given in
table~\ref{tab:inputs}  with GIM and charming
penguins included and taking the value of $\gamma$ determined from the 
UTA, see eq.~(\ref{eq:gamma}). We find
\begin{eqnarray} \vert \hat B_{1} \vert &=& 0.13 \pm 0.02\, , \quad \quad  
 \phi = (188 \pm 82)^o \, , 
\nonumber \\
\vert \hat B^{\sss {\rm GIM}}_{1} \vert &=& 0.17 \pm 0.08\, , \quad \quad  
 \phi^{\sss {\rm GIM}}  = (181 \pm 59)^o \, ,  
\label{eq:fits}
\end{eqnarray}
where the notation is self-explaining.  
Note that the size of the charming  and GIM penguin effects is of the
expected magnitude.    Note that $\vert \hat B^{\sss {\rm 
GIM}}_{1} \vert$ is very  poorly determined. The reason is that  
only $BR(B_{d}^{0} \to \pi^{+} \pi^{-})$ in the fit is sensitive to 
GIM penguins. Thus, in practice, we 
are trying to fit two parameters, namely  $\vert \hat B^{\sss {\rm 
GIM}}_{1} \vert$ and $\phi^{\sss {\rm GIM}}$, to  a single $BR$.

The results for the $BR$s can be found in
table~\ref{tab:one} with the label ``Charming+GIM".  They show that
the extra charming and GIM parameters radically  improve the agreement for the measured $B \to
K \pi $ and $B \to
\pi \pi $ $BR$s.  We do not claim, however, to be able to predict
$BR(B_d \to \pi^+ \pi^-) $ since many effects of the same order besides charming and GIM 
contributions, which in this case are not Cabibbo enhanced,  were ignored: our results instead show that accurate
predictions for $B_{d} \to \pi \pi$ decays can only be obtained by
controlling quantitatively the ${\cal O}(\Lambda_{QCD}/m_{b})$
corrections, which is presently far beyond the theoretical reach.
To give a complete information, and for comparison with ref.~\cite{lorolast}
we also fit the data by letting $\gamma$ free. In this case we obtain 
$\gamma=(89 \pm 42)^o$. At present, the precision of the data and the number
of free parameters does not allow a useful determination of $\gamma$.

 The large absolute values of $\phi$, and the sizable
effects that penguins have on the $BR$s, stimulated us to consider
whether we could find observable particle-antiparticle asymmetries as
the one defined in eq.~(\ref{eq:asy}). We find large effects in
$BR(B^{+} \to K^+ \pi^0) $, $BR(B_d \to K^+ \pi^-) $ and $BR(B_d \to
\pi^+ \pi^-)$.  As discussed before,
for $BR(B_d \to \pi^+ \pi^-)$ our predictions suffer from very large uncertainties due
to contributions which cannot be fixed theoretically. For this reason,
the values of the asymmetry  reported in table~\ref{tab:asy} are only
an indication that a large asymmetry could be observed also in this channel.
There is a  sign ambiguity
 in  ${\cal A}\sim \sin \gamma \sin \phi$. This ambiguity can be
solved only by an experimental measurement or, but this is extremely
remote, by a theoretical calculation of the relevant amplitudes. For
each channel, we give the absolute value of the asymmetry in
table~\ref{tab:asy}.  Note that within factorization all asymmetries
would be unobservably small, since the strong phase is a perturbative
effect of ${\cal O}(\alpha_{s})$~\cite{beneke}.  The possibility of
observing large asymmetries in these decays opens new perspectives. 
These points will be the subject of a future study.
 \begin{table} 
 \centering 
 \begin{tabular}{|c|c|c|c|}  \hline 
 $\vert {\cal A} \vert$ & Charming + GIM &  $\vert {\cal
 A}\vert$ & Charming + GIM  \\  
 \hline 
 $K^0 \pi^0$ & $0.05 \pm 0.03 $& 
$K^+ \pi^0$  &$ 0.16 \pm 0.08$\\ 
 $K^0 \pi^+$ & $ 0.02 \pm 0.02 $& 
$K^+ \pi^-$ & $ 0.15 \pm 0.07$\\ 
 $ \pi^+ \pi^-$ &$ 0.44 \pm 0.21 $& 
 $ \pi^0 \pi^0$ &$ 0.61 \pm 0.29 $\\ 
\hline  
 \end{tabular}  
 \caption{Absolute values of the rate CP asymmetries for $B \to K
 \pi$ and    $B \to \pi \pi$ decays.} 
 \label{tab:asy} 
\end{table} 

\subsection*{Comparison with BBNS}
In this section we make a critical comparison with the  latest analysis of 
BBNS~\cite{lorolast}. We recall that we added all the perturbative corrections
and allowed the variations of the non-perturbative parameters which
 were implemented by BBNS. 

\begin{figure}
\centering
\epsfig{figure=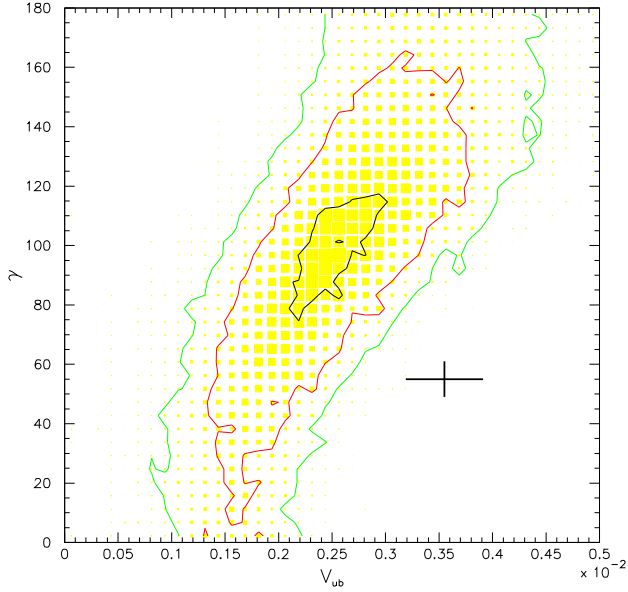,height=3.5in,width=3.5in} 
\caption{Contour plot in the $\gamma - \vert V_{ub}\vert$ plane, see text.}
\label{fig:corr}
\end{figure}

We start by discussing the r\^ole of  chirally enhanced terms that do not
suffer from infrared divergences at  ${\cal O}(\alpha_s)$. This is an istructive
case since, if not for other infrared-divergent contributions of the same order,
the inclusion of only  these terms, although not justified,  would still allow   {\it model independent}
predictions, in the sense discussed before.
Thus the question is whether these chirally enhanced terms alone, part of which
have
the same effect as $\tilde P_1$, can describe the data.
Thus we have repeated the  analysis in the UTA case with only factorization and chirally
enhanced infrared-finite terms with the results for the $BR$s given in 
table~\ref{tab:one} with the label ``chirally enhanced''. The combined probability
that all the predicted values for the $K \pi$ channels are within two $\sigma$s
from the experimental numbers is 6\%. If we relax the constraint on $\gamma$
we obtain $\gamma = (127 \pm 20)^o$ with a probability of 0.3\%that $\gamma 
< 80^o$ and of 2\% that $\gamma < 90^o$. The corresponding  $\rho$--$\eta$ 
contour plot is 
given in fig.~\ref{fig:lorolastnostro} (upper--left).  We conclude that this model
(model in the sense of including only a chosen subset of the chirally enhanced
terms) is strongly disfavoured  by the data. The reader may be surprised of the difference 
between 
fig.~\ref{fig:lorolastnostro} and fig.~17 of ref.~\cite{lorolast} (where the 
non-perturbative parameters $X_{A,H}$ were however included). The difference 
is explained by the fact that we used the experimental measurements of
$\vert V_{ub}\vert$.
If, following ref.~\cite{lorolast,beuc},
we let both $\rho$ and $\eta$ free, the $\rho$--$\eta$ contour plot 
changes and becomes that shown in fig.~\ref{fig:lorolastnostro} (upper-right),
which is very similar to fig.~17 of ref.~\cite{lorolast}. 
For completeness, we also give in fig.~\ref{fig:lorolastnostro}
the p.d.f. of $\gamma$ in the two cases, together
with the p.d.f. obtained with the UTA analysis of ref.~\cite{Ciuchini:2000de}.
When the experimental information on $\vert V_{ub}\vert$ is used the two p.d.f.
have no overlap (lower-left).  In the other case, it is 
still possible to find  values of 
$\gamma$ compatible with the UTA analysis (lower-right), at the price of
 a rather low value of 
$\vert V_{ub}\vert$. The situation is well illustrated by fig.~\ref{fig:corr},
where the contour plot  in the $\gamma$--$\vert V_{ub}\vert$ plane of
the joint p.d.f. from  non-leptonic decays is compared with the UTA 
range  for $\gamma$ and the allowed interval for the measured $\vert V_{ub}\vert$,
at 1$\sigma$.  This figure demonstrates that, given the strong correlation
between $\gamma$ and $\vert V_{ub}\vert$, it is crucial  to take into account
the experimental knowledge of $\vert V_{ub}\vert$.
 An important theoretical remark is in order at this point.
The coefficient $a_6^f$ is enhanced by the ${\cal O}(\alpha_s)$ corrections 
and may play the same r\^ole of charming penguins. It is very scaring that
its actual value is strongly affected by the  contribution of the
chromomagnetic operator  computed at tree level~\cite{lorolast}. 
It is very hard to believe that this contribution  which, besides all
general considerations,  is also  non-local, will remain infrared-finite in higher orders and can be  really
evaluated in this way. 
\begin{figure}[t]
\centering
\begin{tabular}{cc}
\epsfig{figure=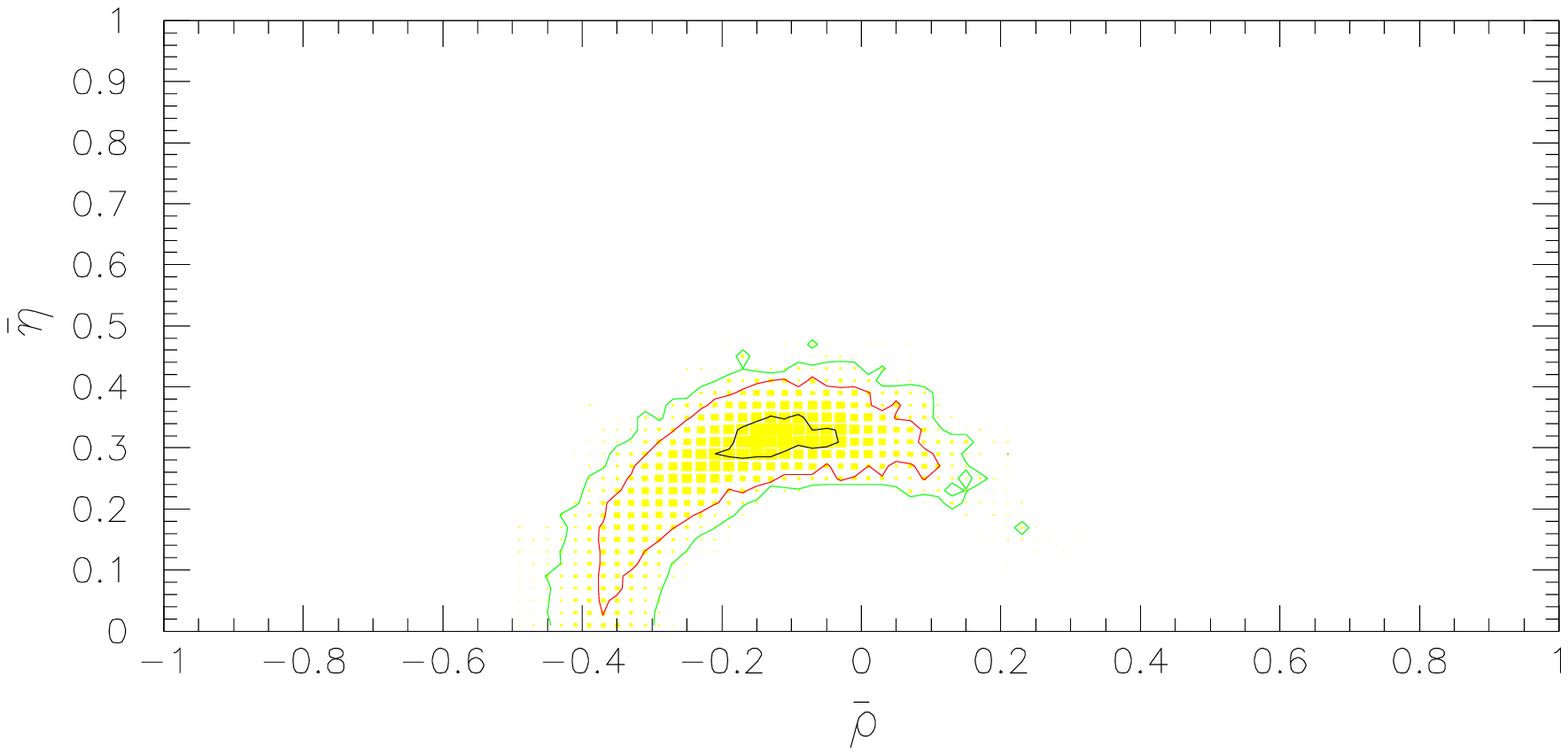,width=3.in} & 
\epsfig{figure=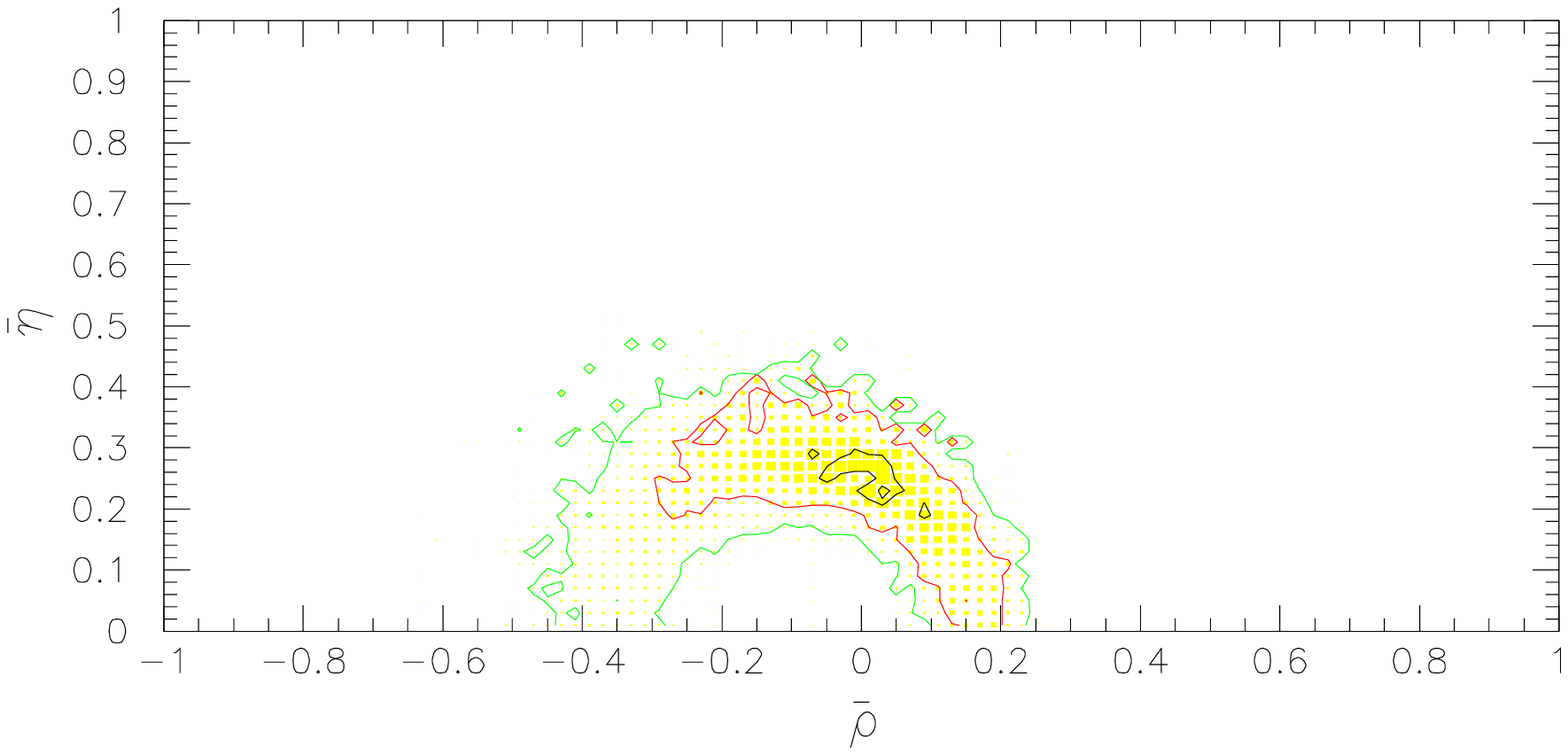,width=3.in} \\ \epsfig{figure=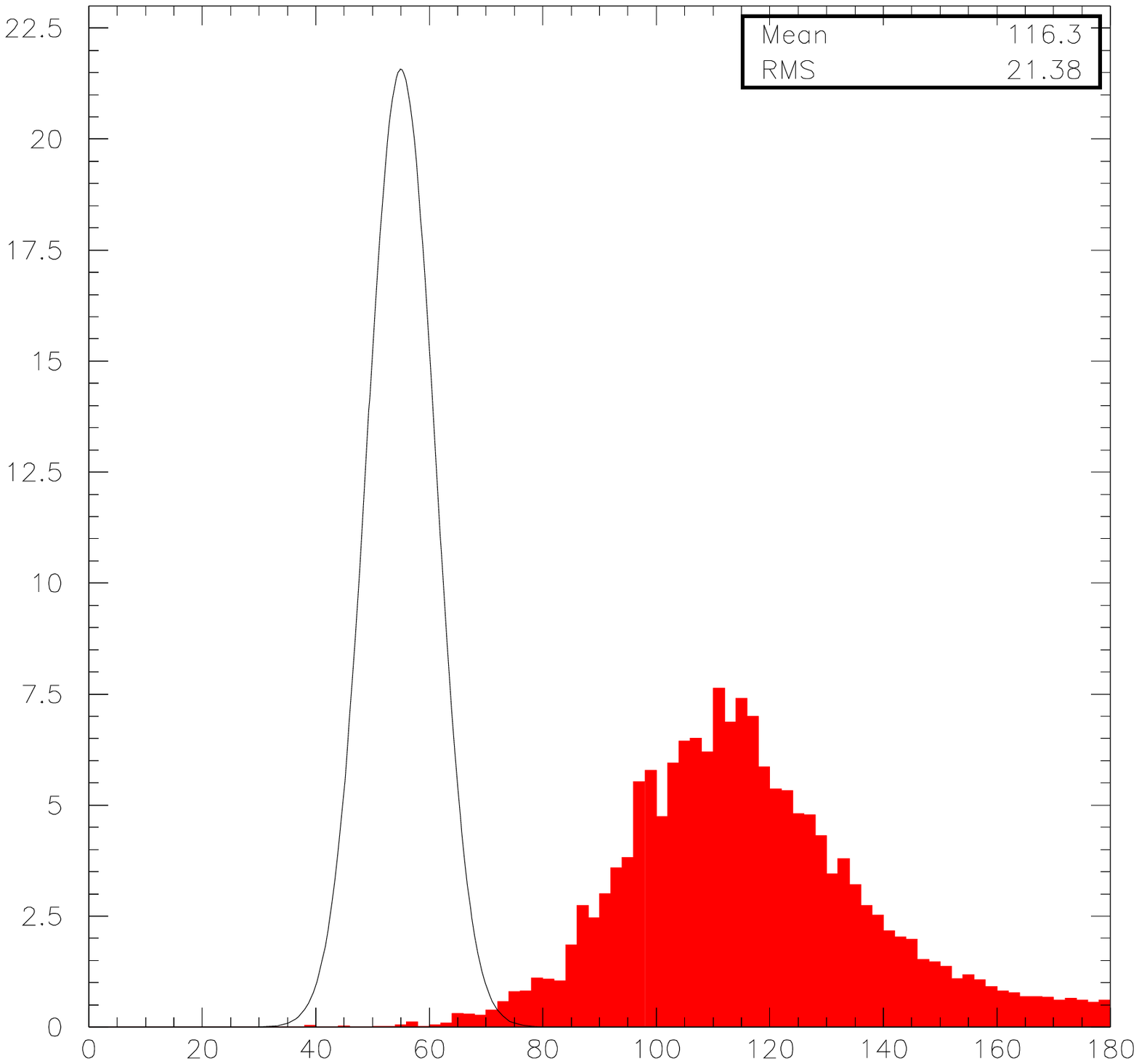,
width=3.in,height=2.8in} & \epsfig{figure=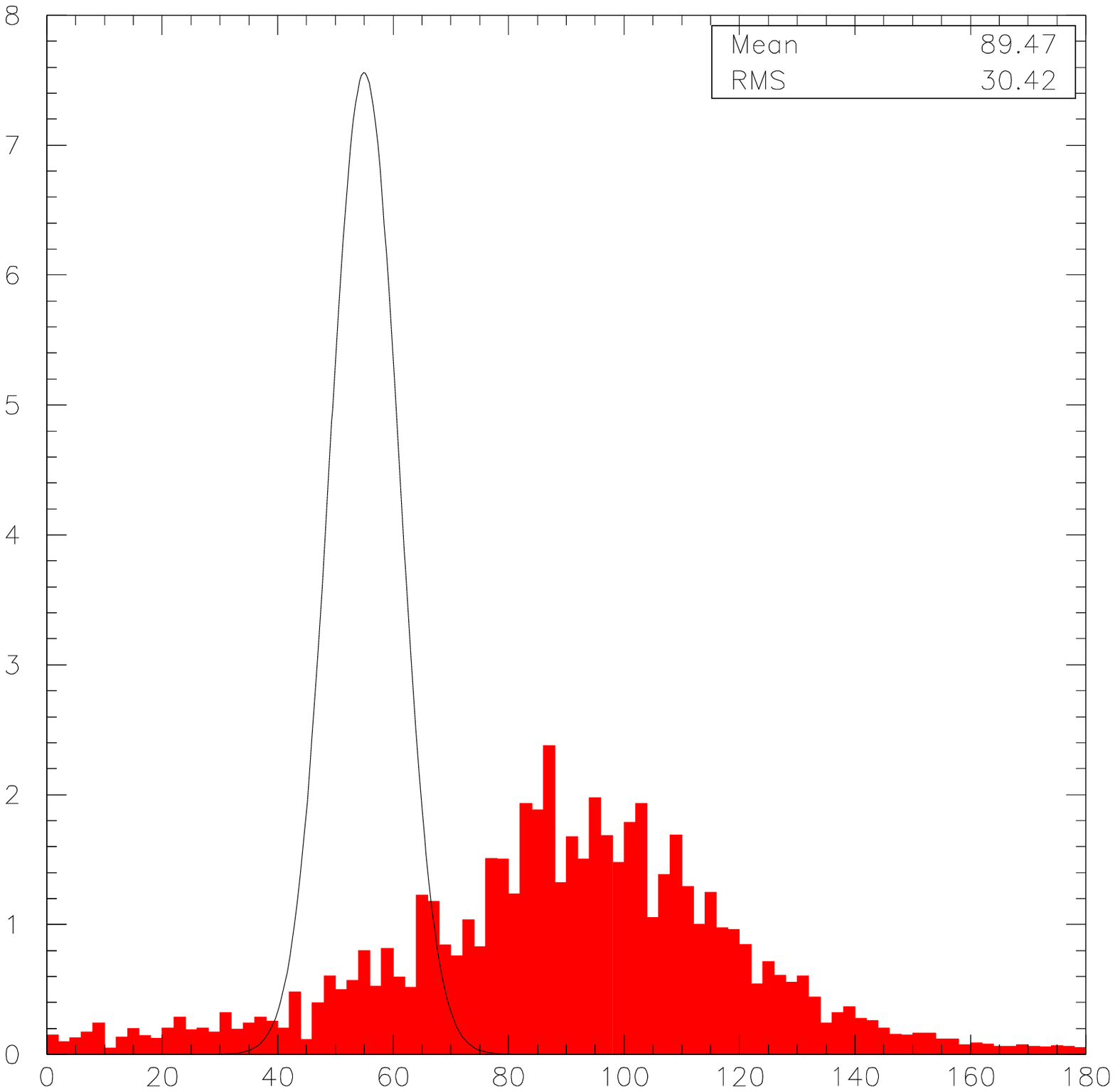,width=3.in,height=2.8in}
\end{tabular}
\caption{$\rho$--$\eta$ contour plots obtained with the BBNS model using $\vert V_{ub}\vert$ 
(up-left) or letting $\rho$ and $\eta$ free (up-right).  
We also show a comparison of the p.d.f. of $\gamma$ with the one from
 the UTA analysis of ref.~\protect\cite{Ciuchini:2000de} 
in the two cases.}
\label{fig:lorolastloro}
\end{figure}

We now discuss the effect of $X_{A,H}$ on the final results. The addition of these 
parameters, in the range proposed in ref.~\cite{lorolast}, leaves the situation
substantially unaltered as shown in fig.~\ref{fig:lorolastloro}. 
This holds true also for the $BR$s, which are given in 
table~\ref{tab:one} with the label BBNS. The reason is that on the one 
hand $X_{A}$ and $X_{H}$ do not have the same quantum numbers, and 
hence effects, of charming penguins, on the other the range chosen 
{\it a priori}, on the basis on one-loop perturbation theory,  is not 
large  enough to improve the agreement with the measured $BR$s (it 
essentially increases the uncertainty on the predictions). Of 
course by choosing a low value of the strange quark mass and of 
$\vert V_{ub} \vert$ (without using the experimental information 
coming from its measurements), a large value of $f_{K}(0)$ and a 
small value for $f_{\pi}(0)$ etc. it is still possible to find some 
point in parameter space where the $\chi^{2}$  is good.  That this can 
be used to fit $\gamma$ and test the SM  is hard to believe though.   

The  situation would be different if $X_{A,H}$ are let free to vary and 
fitted to the data.  We have done this exercise and found that the 
preferred value of $\rho_{A}$ is much larger than the values allowed 
in the interval chosen in \cite{lorolast}, whereas  $\rho_{H}$  is 
not determined by the fit.
We conclude that since $X_{A}$ and $X_{H}$ are infrared divergent 
quantities, the value of which cannot be predicted, and since without 
the inclusion of non-perturbative contributions of ${\cal 
O}(\Lambda_{QCD}/m_{b})$ is not possible to reproduce the experimental 
data, we are bound to use model dependent assumptions in the analysis 
of non-leptonic $B \to K \pi$ and $B \to \pi\pi$ decays. 
\section{Conclusion}
We have analyzed the predictions of QCD factorization for $B \to \pi\pi$
and $B \to K \pi$ decays.  Even taking into account the
uncertainties of the input parameters, we find that QCD factorization is
unable to reproduce the observed $BR$s. The introduction of  charming
and GIM penguins~\cite{charming} allows to reconcile the theoretical
predictions with the data.  Instead of varying the non-perturbative 
phenomenological parameters in preassigned ranges, we prefer to try to 
fit them on the data. 
With the present theoretical and experimental accuracy, we find that it 
is still not possible  to determine the CP violation angle $\gamma$.  The situation 
is expected to improve in the near future with more accurate
experimental measurements of the relevant $BR$s. 

 Contrary to factorization,
we predict large asymmetries for several of the particle--antiparticle
$BR$s, in particular $BR(B^{+} \to K^+ \pi^0) $, $BR(B_d \to K^+
\pi^-) $ and, possibly,  $BR(B_d \to \pi^+ \pi^-)$. This opens new perspectives
for the study of CP violation in $B$ systems.
\section*{Acknowledgments}
We thank  C. Sachrajda for useful discussions on this work.  G.M. and 
L.S. thanks the TH division at CERN where part of this work has been done. 

\end{document}